\documentclass[useAMS,usenatbib]{mn2e}
\usepackage{psfig}
\usepackage{amsmath}
\usepackage{amssymb}
\def \mgii {Mg\,{\sc ii}}
\def \civ {C\,{\sc iv}}
\def \hbeta {H$\beta$}

\title[Stacked reverberation lags at high redshift]{Stacked reverberation mapping}

\author[Fine et al.]
       {S. Fine$^1$\thanks{s.lb.fine@gmail.com},
         T. Shanks$^2$, P. Green$^3$, B. C. Kelly$^3$, S. M. Croom$^4$, R. L. Webster$^5$,
\newauthor
E. Berger$^3$, R. Chornock$^3$, W. S. Burgett$^6$, K. C. Chambers$^6$, N. Kaise$^6$, P. A. Price$^7$\\
$^1$Department of Physics, University of Western Cape, Bellville 7535, Cape Town, South Africa\\
$^2$Department of Physics, Durham University, South Road, Durham DH1 3LE, UK \\
$^3$Harvard-Smithsonian Center for Astrophysics, 60 Garden Street, Cambridge, MA 02138, USA \\
$^4$Sydney Institute for Astronomy, School of Physics, The University of Sydney, NSW 2006, Australia \\
$^5$School of Physics, University of Melbourne, Parkville, VIC 3010, Australia
$^6$Institute for Astronomy, University of Hawaii at Manoa, Honolulu, HI 96822, USA \\
$^7$Department of Astrophysical Sciences, Princeton University, Princeton, NJ 08544, USA\\
}

\begin{document}

\maketitle

\begin{abstract}



Over the past 20\,years reverberation mapping has proved one of the most successful techniques for studying the local ($<1$\,pc) environment of super-massive black holes that drive active galactic nuclei. Key successes of reverberation mapping have been direct black hole mass estimates, the radius--luminosity relation for the \hbeta\ line and the calibration of single-epoch mass estimators commonly employed up to $z\sim7$. However, observing constraints mean that few studies have been successful at $z>0.1$, or for the more-luminous quasars that make up the majority of current spectroscopic samples, or for the rest-frame ultra-violet emission lines available in optical spectra of $z>0.5$ objects.

Previously we described a technique for stacking cross correlations to obtain reverberation mapping results at high $z$. Here we present the first results from a campaign designed for this purpose. We construct stacked cross-correlation functions for the \civ\ and \mgii\ lines and find a clear peak in both. We find the peak in the \mgii\ correlation is at longer lags than \civ\ consistent with previous results at low redshift. For the \civ\ sample we are able to bin by luminosity and find evidence for increasing lags for more-luminous objects. This \civ\ radius--luminosity relation is consistent with previous studies but with a fraction of the observational cost.

\end{abstract}


\section{Introduction}

The inner regions of active galactic nuclei (AGN) offer a unique
opportunity to study matter within a few parsecs of a super-massive black hole. Reverberation mapping is designed to study (primarily) the broad-line region (BLR) of AGN by measuring the interaction between continuum and broad-line flux variations \citep{b+m82,pet93}. The physical model assumes the BLR is photoionised by a UV continuum that is emitted from a much smaller
radius. Variations in the ionising continuum produce correlated
variations in the broad emission-line flux after a delay that can be associated with the light travel time.

Reverberation mapping of a single system requires many spectroscopic epochs of
emission-line and continuum luminosity measurements. The time lag between continuum and emission-line variations is given by the peak in the cross correlation between the two light curves. To date lags have been measured for nearly 50 objects following this approach. 

Reverberation mapping has led to significant advances in the
understanding of AGN (e.g. the radius--luminosity relation;
\citealt{wpm99,ben06}, stratification and kinematics of the BLR
\citealt{p+w99,p+w00}, black hole mass estimates \citealt{pea04}
etc.). However, traditionally these campaigns have been observationally expensive as they require many observations of individual objects over a long period of time. Furthermore, these constraints have meant that even at moderate redshift ($z>0.4$) there are no reverberation lags measured with the exception of one tentative result at $z=2.2$ \citet{kas07}. This also means that, apart from some notable exceptions that have had multi-epoch UV spectroscopy, the vast majority of measured lags are for the \hbeta\ line.

In a previous paper \citep{me6} we discussed the potential gains of combining extensively-sampled time-resolved photometric surveys, specifically the Pan-STARRS1 medium-deep survey (PS1 MDS), with relatively few epochs of spectroscopy on large samples of QSOs. Current multi-object spectrographs make it possible to rapidly obtain spectra of many hundreds of QSOs. While individual objects may not have a well-sampled emission-line light curves \citet{me6} showed (in simulations at least) that reverberation lags could be recovered through stacking.

In this paper we present the first results from a multi-epoch spectroscopic survey of the PS1 MDS fields with the aim of measuring reverberation lags in stacked QSO samples. Throughout we assume $H_0,\,\Omega_M,\,\Omega_\lambda=70,\,0.3,\,0.7$.


\section{Data}

In \citet{me6} we discussed the sources of our data and their reduction. We only repeat the main points briefly here. See \citet{me6} for a more detailed description.

\subsection{PS1 light curves}

The PS1 telescope \citep{hod04} is performing a series of photometric surveys of the northern sky. The MDS offers the best opportunity for our analysis. Images of the ten MDS fields are taken every four-five nights in the $grizy$-bands while not affected by the sun or moon \citep{kai10,ton12}.
We use {\sc psphot}, part of the standard PS1 Image Processing Pipeline system \citep{mag06}, to extract point-spread-function photometry from
nightly stacked images of the MDS fields. Each nightly stack is
divided into $\sim70$ skycells. We calibrate each skycell separately
using SDSS photometry \citep{fuk96,yor00} of moderately bright
($16<\text{Mag.}<18.5$) point sources ($type=6$ in the SDSS database). Each of the light curves is inspected for outliers that are removed manually, in total $<1$\,\% of the data points were removed.


\subsection{Hectospectra}

QSOs in the MDS fields are being observed for an ongoing project to study the variability of QSOs. QSO candidates are selected for spectroscopy using photometric databases of QSOs \citep{ric09,bov11} to which we add point sources that correspond to X-ray sources, variability selected objects and UVX selected objects.

We are surveying the MDS fields with the Hectospec instrument
on the MMT. Each MDS field is tiled with seven MMT
pointings. Exposures are $\sim$1.5\,h in length meaning that an MDS field
($\sim500$ QSOs) can be surveyed in $\sim1$\,night of good-quality on-sky observing time. The spectra are extracted and reduced using standard Hectospec
pipelines \citep{min07}. They are then flux calibrated using observations of F~stars in the same fields.

Spectra are classified and redshifted manually using the {\sc runz} code \citep{col01,dri10}. So far from 6.5 nights awarded we have observed 15 Hectospec fields (5 more than once). Further spectra were obtained from spare fiber allocations by the Pan-STARRS transient group. In total we have spectra of 2727 objects, 1228 of which are QSOs, 368 of which have $>1$ spectroscopic epoch.

\section{Analysis}

\subsection{K-corrections}

From the MDS we have $griz$-band light curves for each object in our sample. We K-correct these magnitudes using a simple model fit to the SDSS $ugriz$-band magnitudes. We fit a powerlaw along with a template for QSO emission lines and a Lyman-$\alpha$ break (see e.g. \citealt{croom09b} for a similar if more detailed approach). While this model is not always a good fit to the SDSS magnitudes, it suffices for the purposes of interpolating K-corrections for this work.

We K-correct all of our light curves to a single rest wavelength (1350 and 3000\,\AA\ for \civ\ and \mgii\ respectively). Note that the value of this K-correction affects any cross-covariance analysis. On the other hand cross-correlations are normalised and hence unaffected by the value of the K-correction (see section~\ref{sec:xc_anal}).

\subsection{Line flux measurements}

\begin{figure}
\centerline{\psfig{file=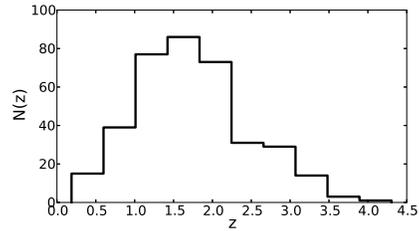,width=6.cm}}
\caption{The redshift distribution of the 368 QSOs in our sample with more than one spectrum.}
\label{fig:nz}
\end{figure}

In Fig.~\ref{fig:nz} we show the redshift distribution of the 368 QSOs
with $>1$\,spectrum. The distribution is relatively
typical of optically selected QSO surveys with the majority between $z\sim0.5$ and 3. Over this redshift range the strongest, and most heavily studied, broad emission lines are \mgii\ ($\lambda 2789$) and \civ ($\lambda 1549$). We will focus on these two lines throughout the rest of this work.

We fit the \mgii\ and \civ\ lines following the multiple-Gaussian prescription outlined in \citet{me2,me3} and each fit is manually inspected to check the reliability. Note that the standard Hectospec pipeline does not return a variance array for the spectra. However, from calculating the scatter in flat regions of the spectra it is clear that the spectral S/N is $\gtrsim10$\,pix$^{-0.5}$ for the majority of sources. Errors on the emission-line fluxes are therefore dominated by flux calibration errors that we find to be $\sim5-10$\,\%\ by comparing the fiber magnitudes of calibration F stars with their SDSS photometry. Note that this will introduce correlated errors into our flux measurements since a single flux calibration is used for each Hectospec field.

Manual inspection was performed to remove bad fits, sky residuals, broad and narrow-line absorption (in instances where narrow absorption lines or sky residuals did not significantly effect the emission line the effected pixels were simply masked and the line was fitted as normal) and other potential sources of contamination. The manual inspection was carried out several times with varying degrees of exclusivity. In general we found that more exclusive selection (i.e. more objects thrown out) produced he highest fidelity in our results. After this procedure we were left with 89 (\mgii) and 75 (\civ) objects with a good fit to their emission lines at more than one spectroscopic epoch.

\subsection{Cross correlation analysis}
\label{sec:xc_anal}

The correlation coefficient $r$ for two samples ($x$, $y$) is
\begin{equation}
r(x,y)=
\frac{\text{Cov}(x,y)}{\sigma_x\sigma_y}=
\displaystyle\sum\limits_{\substack{i,j}} \frac{(x_i-\overline{x})(y_j-\overline{y})}{\sigma_x\sigma_y}
\label{equ:cf}
\end{equation}
where Cov$(x,y)$ is the covariance between the samples and $\sigma$ is the rms of the samples. \citet{me6} concentrated on cross covariances rather than cross correlations primarily because the rms of the emission line fluxes is poorly constrained from just two measurements. In practice we find it considerably more favorable to use cross-correlations in our analysis. We do this since QSO-to-QSO variations require normalisation of the covariance function, and while the rms may not always be well defined it does improve the quality of the stacked results. Nevertheless, most of the results presented here for cross correlations are visible, if with decreased clarity, in the covariance functions as well. Furthermore, cross correlations are not susceptible to the step-biases we found in cross covariances in \citet{me6}.

To calculate the stacked cross correlations we take each emission-line flux/continuum photometric observation pair for each object in our sample. We calculate the time lag $\tau$ between them and the individual $i,j$th term of equation~\ref{equ:cf}. We bin all of the emission line-continuum data-data pairs by their lag and average over all pairs within a bin to create the stack.

In practice we tried mean, variance weighted and median stacks. Each of these methods produced roughly equivalent results for the \civ\ stack. For \mgii\ we found variance weighting gave the highest S/N in the cross correlation. Given the sparsity of solid reverberation-mapping results at $z>0.3$, this paper's primary goal is to identify a(ny) peak in the stacked correlation functions presented here. The variance weighted mean biases the results towards the higher quality data (in practice the brighter objects) and has the effect of reducing the noise in the final stack. Because we find it gives the clearest results for \mgii\ we will use the variance weighted mean to create our stacked correlation functions throughout this paper.



To estimate errors on the stacks we use the field-to-field technique whereby we divided our sample into nine subsamples. We then calculated the stacked cross-correlation functions in each subsample and took the rms of the subsamples divided by $\sqrt{9}=3$ as an estimate of the error.

\subsection{Randomised analysis}
\label{sec:ran_anal}

As a check on the significance of any results we obtain we also performed a randomised analysis. For this we take the spectroscopic flux measurements for each individual object and replace them with random values drawn from a Gaussian distribution with the same mean and rms as the observations (note that in most cases the mean and rms are only defined by two observations). We then perform the same stacked cross-correlation analysis with the observed photometric light curves. This process was repeated 200 times and the mean and rms of the separate realisations was recorded for comparison with our results.

\section{Results}


\subsection{The full \civ\ stack}

\begin{figure}
\centerline{\psfig{file=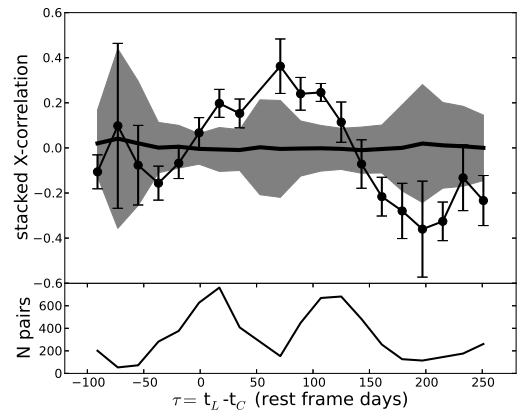,width=8.cm}}
\caption{The stacked \civ\ cross correlation for all objects with $>1$ good emission line measurement. The thick line and shaded region shows the mean and rms of the random simulations we performed. In the bottom panel we show the number of continuum/emission line flux pairs that go into the stack at each point (i.e. the number summed over in equation~\ref{equ:cf}).}
\label{fig:xc_civ1}
\end{figure}

Figure~\ref{fig:xc_civ1} shows the variance-weighted stacked cross-correlation function for all 75 objects with more than one epoch of \civ\ data. The lower panel in the plot shows the number of data--data pairs that contribute to the stack in each $\tau$ bin. Since PS1 surveys the MD07 field for $\sim6$\,months of each year we sample lags of $\pm0.5$\,integer~years poorly. Once transformed to rest-frame time delays this creates the peaks in the N pairs distribution.

Each object in the \civ\ sample will have a different lag for its \civ\ line due to a variety of reasons, perhaps most importantly their different luminosities. The \civ\ sample spans $37.5<\log_{10}\lambda L_\lambda(1350)<39.5$ assuming $\tau\propto L^{0.5}$ this gives a potential factor-of-10 range in lags within our sample. The effect of stacking objects with a broad range of lags would be to smooth our results. Despite this potential smoothing we find a significant peak in the cross-correlation function in Figure~\ref{fig:xc_civ1}. The solid line and shaded area in the figure show the mean and rms from our 200 simulated cross correlations where we randomise the spectroscopic flux measurements (section~\ref{sec:ran_anal}). Seven of our cross correlation points lie above the rms level (and five below) indicating a significant correlation between our photometric and spectroscopic datasets.

\subsection{Towards a radius--luminosity relation for \civ}

\begin{figure}
\centerline{\psfig{file=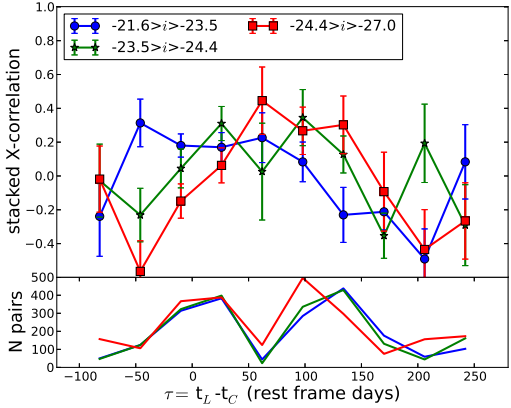,width=8.cm}}
\caption{The stacked \civ\ cross correlation for three magnitude bins with equal numbers of objects.}
\label{fig:xc_civ3}
\end{figure}

To reduce the smoothing caused by the range of lags in each stack, and to aim towards a more useful result to the community (measuring the $r$--$L$ relation for \civ), we bin the sample by the $i$-band absolute magnitude. Figure~\ref{fig:xc_civ3} shows the stacked cross correlation for three equal size (in terms of number of objects) magnitude bins.

\begin{figure}
\centerline{\psfig{file=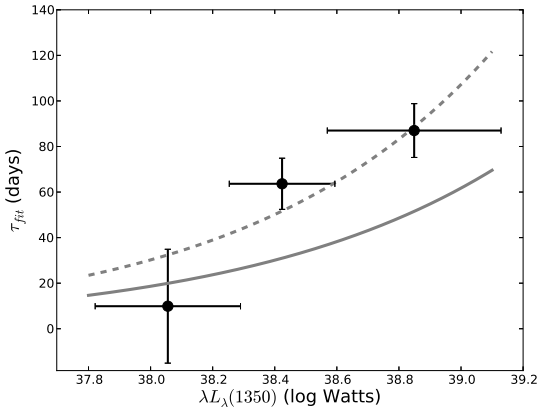,width=8.cm}}
\caption{The $r$--$L$ relation for the \civ\ line where $\tau_{fit}$ is derived from Gaussian fits to the data in Fig.~\ref{fig:xc_civ3}. The grey solid and dashed lines give the \citet{kas07} FITEXY and BCES fits to their $r$--$L$ relation respectively.}
\label{fig:civ_rl}
\end{figure}

There is some evidence for a peak in the two brighter bins, and perhaps in the faintest as well. There is also a suggestion that the peak is shifting towards longer lags in the more-luminous bins. To quantify this we fitted offset Gaussian functions to each of the cross correlations. In figure~\ref{fig:civ_rl} we plot the centroid of the fitted Gaussian against the mean continuum luminosity at 1460\AA\ for each bin. Errorbars in figure~\ref{fig:civ_rl} come from the Gaussian fit for $\tau_{fit}$ and the rms of the continuum luminosities for $\lambda L_\lambda$. The grey lines in the figure are the two fits \citet{kas07} present for their \civ\ $r$--$L$ relation.

There are competing biases that effect the location of the peak in our stacked cross correlations. For example brighter objects tend to have higher S/N photometry and spectroscopy giving them more weight in the variance weighted mean stacks. On the other hand fainter QSOs tend to be more variable potentially increasing their impact on the stacks. Although more data is clearly needed to determine its exact form, there is tantalising evidence for the existence of a \civ\ radius luminosity relation.



\subsection{The \mgii\ correlation}

\begin{figure}
\centerline{\psfig{file=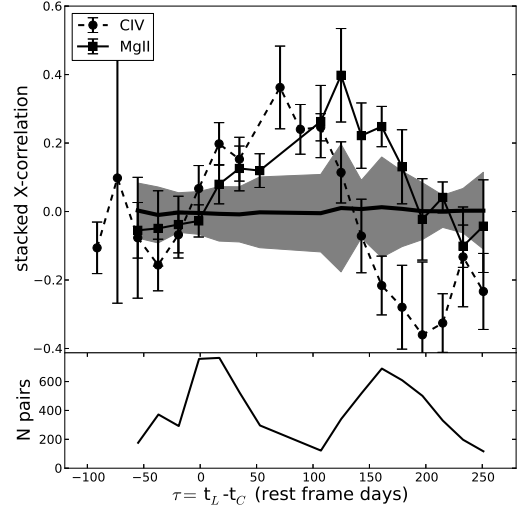,width=8.cm}}
\caption{\emph{Solid:} The stacked cross correlations for the whole \mgii\ sample. The dashed line shows the \civ\ results and the heavy line and shaded region show the mean and rms from our random simulations (section~\ref{sec:ran_anal}).}
\label{fig:xc_mgc}
\end{figure}

While the \mgii\ sample is slightly larger than \civ\ we do not find this translates to a more signal in the stacked correlation function. However, in the full variance-weighted stack we do find a clear peak. The stacked correlation for the \mgii\ sample is given in figure~\ref{fig:xc_mgc} along with the \civ\ stack for comparison. The heavy line and shaded area in the figure show the mean and rms from our 200 simulated cross correlations where we randomise the \mgii\ flux measurements. Again, a significant peak is evident in the \mgii\ cross correlation.

The location of the \mgii\ peak is at considerably longer lags than for \civ. This is consistent with results for low redshift objects where higher ionisation lines such as \civ\ exhibit shorter lags. We note however, that in our sample the situation is complicated by the fact that the \mgii\ sample is by necessity lower redshift than the \civ, hence the objects contributing to the stack tend to be less-luminous which will bias the \mgii\ lag with respect to the high-$z$, luminous \civ\ sample. We have tried splitting up the \mgii\ sample into magnitude bins to investigate any $r$--$L$ relationship but find no convincing peaks when binned.

\section{Discussion}

Reverberation mapping results are the basis for a large part of our understanding of AGN. With some notable exceptions the vast majority of SMBH mass estimates for AGN come directly or indirectly from reverberation mapping. In particular single-epoch estimators \citep{vest02,m+j02}, that have been applied to $>100,000$ objects have their basis in reverberation mapping, rely on the $r$--$L$ relation and are calibrated against reverberation masses.


As of yet there is no published $r$--$L$ relation for the \mgii\ line. \citet{kas07} present a relation for the \civ\ line based on the six objects that have had a reverberation lag measured for them. However, of these objects four have very similar luminosities and cannot be used to define the gradient of the relation. Their gradient is defined mostly by their results for S5~0836+71, the only high-redshift QSO to have a lag measured, and the dwarf Seyfert NGC~4395, an extremely under-luminous AGN \citep{pet05}.

In this letter we present the first results from our campaign to derive reverberation signals from stacks of objects. The technique naturally lends itself to the mapping of rest-frame ultra-violet lines (\mgii\ and \civ) at high redshift. We show that from the 75 objects in our \civ\ stack we get a clear peak in the cross correlation. When binned by luminosity there may be some evidence for a $r$--$L$ relation although our data are not well constrained. However, our data are in good agreement with the \civ\ $r$--$L$ relation from \citet{pet05} and \citep{kas07}. While still at an early stage, the consistency of the Peterson results at $z\sim0$, ours at $z\sim2$, and Kaspi's at $z=2.2$ appears to indicate little evolution in the $r$--$L$ relation with redshift.

We also find a peak in the stacked \mgii\ sample. Comparing with the \civ\ results we find the \mgii\ peak to be at larger $\tau$, indicative of a stratified BLR as observed in low-redshift objects. However, our \mgii\ and \civ\ samples are at different redshifts, with different luminosity distributions confusing a direct comparison.


\section{Conclusions}

We have shown that by stacking samples of QSOs that have continuous photometric monitoring and two-or-more spectra it is possible to recover a reverberation-mapping time lag. We give stacked cross correlations for both the \mgii\ and \civ\ lines and find a clear peak in both. The \mgii\ peak is at considerably longer lags indicative of stratification of the BLR. Further more when binned by luminosity the \civ\ sample shows evidence for increasing lags with increasing luminosity. From these data we make an initial $r$--$L$ plot for the \civ\ line. Although we caution that our relation is effected by significant biases we find it is consistent with previous evaluations.

This paper demonstrates the potential of the stacking technique to produce reverberation-mapping results at high redshift. Potentially this technique could provide an avenue towards answering some key questions about high-redshift AGN. However, it is clear that more data are required before strong constraints can be derived from this technique.

\section{Acknowledgments}

SF would like to acknowledge SKA South Africa and the NRF for their funding support. The data presented in this work came from the Pan-STARRS1 telescope and the Multiple Mirror Telescope.
Observations reported here were obtained at the MMT Observatory, a joint facility of the Smithsonian Institution and the University of Arizona.
The Pan-STARRS1 Surveys (PS1) have been made possible through contributions of the Institute for Astronomy, the University of Hawaii, the Pan-STARRS Project Office, the Max-Planck Society and its participating institutes, the Max Planck Institute for Astronomy, Heidelberg and the Max Planck Institute for Extraterrestrial Physics, Garching, The Johns Hopkins University, Durham University, the University of Edinburgh, Queen's University Belfast, the Harvard-Smithsonian Center for Astrophysics, the Las Cumbres Observatory Global Telescope Network Incorporated, the National Central University of Taiwan, the Space Telescope Science Institute, and the National Aeronautics and Space Administration under Grant No. NNX08AR22G issued through the Planetary Science Division of the NASA Science Mission Directorate.

\bibliography{bib}

\end{document}